\pdfoutput=1

\documentclass[11pt]{article}

\usepackage[]{acl}

\usepackage{times}
\usepackage{latexsym}

\usepackage[T1]{fontenc}

\usepackage[utf8]{inputenc}

\usepackage{microtype}

\usepackage{inconsolata}
\usepackage{booktabs}
\usepackage{xspace}
\usepackage{amsmath}
\usepackage{amssymb}
\usepackage{mathtools}
\usepackage{tcolorbox}
\usepackage{xspace}
\usepackage{caption}
\usepackage{subcaption}
\usepackage[hang,flushmargin]{footmisc}

\newcommand{\LRL}[0]{LRL\xspace}
\newcommand\mrtydi{Mr.~T{\small Y}D{\small I}\xspace}

\title{Zero-Shot Listwise Document Reranking with a Large Language Model}

\author{Xueguang Ma, Xinyu Zhang, Ronak Pradeep, Jimmy Lin \\[1ex]
David R. Cheriton School of Computer Science,\\
University of Waterloo, Canada \\[1ex]
\texttt{\{x93ma, x978zhang, rpradeep, jimmylin\}@uwaterloo.ca} \\}

\begin{document}
\maketitle
\begin{abstract}

Supervised ranking methods based on bi-encoder or cross-encoder architectures have shown success in multi-stage text ranking tasks, but they require large amounts of relevance judgments as training data.
In this work, we propose \textbf{L}istwise \textbf{R}eranker with a \textbf{L}arge Language Model (LRL), which achieves strong reranking effectiveness without using any task-specific training data.
Different from the existing \textit{pointwise} ranking methods, where documents are scored independently and ranked according to the scores, 
LRL directly generates a reordered list of document identifiers given the candidate documents.
Experiments on three TREC web search datasets demonstrate that \LRL not only outperforms zero-shot pointwise methods when reranking first-stage retrieval results, but can also act as a final-stage reranker to improve the top-ranked results of a pointwise method for improved efficiency.
Additionally, we apply our approach to subsets of MIRACL, a recent multilingual retrieval dataset, with results showing its potential to generalize across different languages.\footnote{This work was conducted in late 2022 and was rejected from ACL 2023.}
\end{abstract}

\section{Introduction}

Text retrieval is a critical upstream task for many open-domain natural language understanding capabilities such as question answering and fact verification.
It involves finding and ranking the most relevant documents or text snippets in response to a given query~\citep{drqa, kilt}.
Supervised retrievers and rerankers have shown success in forming multi-stage ranking pipelines for text ranking across multiple domains and languages~\citep{monobert, beir, mmsarco}.
However, these methods rely on the availability of large amounts of labeled data, such as the MS MARCO dataset~\citep{msmarco}, to train the models.
One remaining open question is how to build a text ranking pipeline that can perform well without using any human relevance judgments. 

\begin{figure}
    \centering
    \begin{subfigure}[b]{0.45\textwidth}
    \centering
    \includegraphics[width=\textwidth, trim={0 14em 0 0},clip]{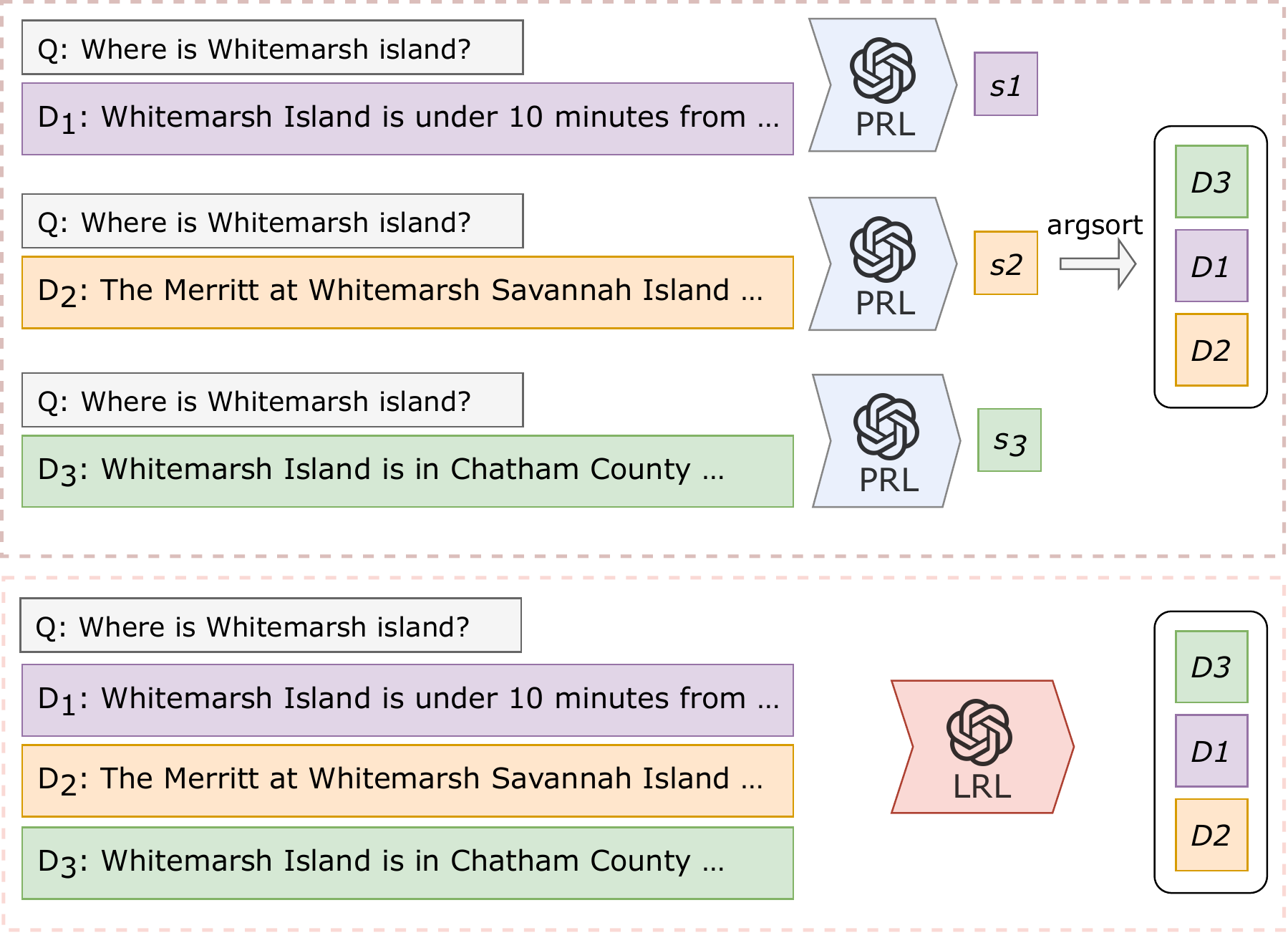}
    \caption{Pointwise reranking pipeline.}
    \label{fig:prl}
    \end{subfigure}

    \begin{subfigure}[b]{0.45\textwidth}
    \centering
    \includegraphics[width=\textwidth, trim={0 0 0 21em},clip]{LRL.drawio.pdf}
    \caption{Listwise reranking pipeline.}
    \label{fig:lrl}
    \end{subfigure}
\caption{Comparison between pointwise and listwise reranking pipelines.}
\label{fig:demo}
\end{figure}

In the current multi-stage ranking paradigm, the reranker usually takes a query--document pair as input and outputs a relevance score. 
Typically, documents from a list of candidates are processed independently without information about each other.
Previous work in supervised reranking tackles this shortcoming in the training process by updating the models based on pairwise or listwise loss (e.g., contrastive learning), which is computed from a pair or a list of candidate documents associated with the same query~\cite{lce, rankt5}.
Doing so forces the models to learn the relative relevance between documents.
However, a zero-shot reranker without any task-specific training cannot take advantage of such an approach.

Inspired by the strong zero-shot effectiveness of large language models~\cite{InstructGPT, T0, FLAN}, we propose Zero-Shot \textbf{L}istwise \textbf{R}eranker with a \textbf{L}arge Language Model (LRL).
Specifically, we use GPT-3 as a zero-shot listwise reranker that takes a query and a list of candidate documents as input and directly generates a reordered list of document identifiers based on the relevance of the documents to the query.
We hypothesize that by taking multiple documents into account, a zero-shot listwise reranker can be more effective than a pointwise one.
\autoref{fig:demo} contrasts the pointwise and listwise pipelines.

We conduct experiments to evaluate the effectiveness of \LRL on a variety of datasets.
These include the TREC DL shared tasks from three years (DL19, DL20, and DL21)~\citep{dl19, dl20, dl21}, as well as three non-English datasets from MIRACL~\cite{miracl}.
Our results show that \LRL outperforms pointwise rerankers on the TREC DL datasets by around 6 points nDCG@10 on average over UPR, the current best-known zero-shot reranker.
\LRL also outperforms BM25, achieving an average improvement of 15 points nDCG@10 on a subset of MIRACL~\cite{miracl}, a recent multilingual text retrieval dataset.
While there are concerns about the mingling of training and test data within large language models, these results should be interpreted with the following mitigating factors:\
First, our central claim is about pointwise vs.\ pairwise methods, which is orthogonal to ``dataset pollution'' concerns.
Second, MIRACL is a relatively new dataset and hence less likely to suffer from these issues (yet).

\section{Background and Related Work}

Given a corpus $\mathcal{C}=\{D_1, D_2, ..., D_n\}$ that contains a collection of documents and a query $q$, the task of a \textit{retriever} is to return a list of $k$ documents from $\mathcal{C}$ that are most relevant to query $q$ according to some metric, where $k\ll |\mathcal{C}|$.
The task of a \textit{reranker} is to further improve the quality of the ranked list according to either the same or a different metric.
Retrievers and rerankers together form multi-stage ranking pipelines for text ranking~\cite{duobert, lce}.
Despite recent work on supervised neural retrievers that have shown much success in text ranking tasks, they all require a large amount of task-specific human-labeled query--document pairs to train a supervised model.
In this work, we focus on the zero-shot setting, where the model is not provided any task-specific supervised training.

BM25~\citep{bm25} is a traditional lexical retriever based on the probabilistic relevance model, which estimates query--document relevance according to statistics such as term frequency and inverse document frequency.
It does not require manually labeled data and thus fits zero-shot constraints naturally.
Contriever~\citep{contriever} is an unsupervised dense retriever trained by momentum contrastive learning using pseudo query--document pairs cropped from raw texts.
It also does not require task-specific supervision.

On top of BM25 and Contriever, UPR~\citep{upr} uses a zero-shot question generation model to rerank retrieved documents based on query likelihood.
Concretely, UPR estimates the log-probability of whether document $D$ is relevant to query $q$ using a pretrained language model to compute the average log-likelihood of the query tokens conditioned on the document.
As of our efforts, \{BM25, Contriever\} + UPR reranking forms the state-of-the-art (SoTA) zero-shot multi-stage ranking pipeline~\citep{upr}.

Recent work such as InPars~\citep{inpars, inpars-light} and Promptagator~\citep{promptagator} explores using LLMs to generate synthetic queries for documents to craft positive query--document pairs as training data for retrievers or rerankers.
Similarly, HyDE~\citep{hyde} uses LLMs to augment queries by generating hypothetical documents for unsupervised dense retrieval.
In this work, we use LLMs directly as a reranker in a multi-stage ranking pipeline.

\section{Methods}

\subsection{Listwise vs.\ Pointwise Rerankers}

Given a user query $q$ and candidate documents $\{D_1, 
\ldots, D_n\}$ from a corpus or the previous ranking stage,
a \textit{pointwise} reranker (e.g., UPR) takes the query $q$ and a single document $D_i$ and outputs a score $s_{q, D_i}$ indicating query--document relevance.
The candidate documents are then reranked according to these scores. 
This approach holds even when the reranker is based on a sequence-to-sequence (seq-to-seq) model.
Take monoT5~\cite{monoT5} as an example:\ it is based on the seq-to-seq T5~\cite{t5} backbone, where the model is asked to generate the \texttt{True} or \texttt{False} token as output.
The reranker is optimized on the logits of the two tokens as a classification task and outputs a score for each query--document pair. 

In contrast to pointwise rerankers, we propose LRL, a zero-shot \textit{listwise} reranker that leverages large language models.
Specifically, \LRL takes the query $q$ and a list of documents $\{D_1, ..., D_m\}$ as input and returns a reordered list of the input document identifiers.
The key difference is that LRL considers information from multiple documents simultaneously while reranking.
In the context of transformers, this means the model can attend to all the candidate documents to determine their relative ranking position.
Moreover, the listwise reranker directly outputs an ordered list (of document identifiers), which forms a ``seq-to-seq'' process and is naturally suited to large language models.
The comparison between pointwise and listwise reranking pipelines can be seen in \autoref{fig:demo}.

\subsection{Prompt Design}
\label{sec:prompt}

We use a simple instruction template to prompt the LLM to perform zero-shot listwise ranking:

\begin{small}
\begin{verbatim}
Passage1 = {passage_1}
...
Passage10 = {passage_10}
Query = {query}
Passages = [Passage1, ..., Passage10]
Sort the Passages by their relevance to the Query.
Sorted Passages = [
\end{verbatim}
\end{small}

\noindent 
Candidate passages are listed on each line and followed by the target query and an instruction.
The square bracket at the end is left unclosed intentionally to encourage the model to complete \texttt{Sorted Passages} with a list of passage IDs, where the first item is expected to be the most relevant.

For a fair comparison of listwise and pointwise reranking, we also craft a variant of the prompt for zero-shot pointwise reranking.
We instruct the model to predict the relevance of a query--passage pair as follows:

\begin{small}
\begin{verbatim}
Passage: {passage}
Query: {query}
Is this passage relevant to the query? 
Please answer True/False.
Answer:
\end{verbatim}
\end{small}

\noindent This makes the probability distribution of the first decoded token focused on the tokens \texttt{True} and \texttt{False}, and we use the probability of decoding the token \texttt{True} as the relevance score.
We denote this baseline as PRL (\textbf{P}ointwise \textbf{R}eranker with a \textbf{L}LM).

\subsection{Progressive Reranking}
\label{sec:rerank}
\label{longlist}

Due to the constraints of runtime and memory cost, transformer-based models enforce input length limits; for example, GPT-3 limits the maximum input length to 4,000 tokens.
Our listwise reranker may be tasked with reranking a list of documents that is longer than the number of documents that the model can handle.
We propose to solve this issue with a sliding window strategy.

Concretely, assuming the model is able to rerank $m$ documents at a time, in order to rank a list with length $n>m$, we start with ranking the last $m$ documents from the list.
Then we shift the ranking window towards the head of the list by $m/2$ and rerank.
This operation is repeated until the window reaches the head of the list.
Ideally, this strategy should be able to ``promote'' the $m/2$ most relevant documents to the head of the list in one round.
Note that this strategy is not able to completely reorder the entire candidates list, but we demonstrate its effectiveness in improving top-ranked results.

\section{Experimental Setup}

To demonstrate the effectiveness of LRL, we compare it with existing representative unsupervised ranking pipelines:\ BM25, Contriever, UPR, as well as our baseline PRL.

\begin{table*}[t]
\small
\centering
\resizebox{1\textwidth}{!}{ 
\begin{tabular}{l|cc|cc|cc|cc}
\toprule
& \multicolumn{2}{c|}{\textbf{Source}} & \multicolumn{2}{c|}{\bf DL19} & \multicolumn{2}{c|}{\bf DL20}   & \multicolumn{2}{c}{\bf DL21}  \\
&  \textbf{prev.} & \textbf{top-}$k$ & \textbf{nDCG@10} & \textbf{MRR@10} & \textbf{nDCG@10} & \textbf{MRR@10} & \textbf{nDCG@10} & \textbf{MRR@10} \\
\midrule
    \multicolumn{9}{c}{\textit{Zero-shot}} \\
(1) BM25             & None & |$\mathcal{C}$|            & 0.5058  & 0.7024 & 0.4796  &  0.6533 & 0.4458  & 0.4981  \\
(2) Contriever       & None & |$\mathcal{C}$|            & 0.4454  & 0.5928 & 0.4213  &  0.5408  &    --     &  --  \\
\midrule
(3) UPR & BM25 & 100 & 0.5910 & 0.6494 & 0.5958 & 0.7247 & 0.5621 & 0.6956 \\
(4) PRL & BM25 & 100 & 0.5975 & 0.7347 & 0.6088 & 0.7699 & 0.5678 & 0.7148 \\
(5) \LRL & BM25 & 100 & 0.6580 & 0.8517 & 0.6224 & 0.8230 & 0.5996 & \textbf{0.8113} \\
\midrule
(6) \LRL & UPR  & 10 & 0.6382 & 0.8320 & 0.6357 & \textbf{0.8256} & 0.5867 & 0.7543 \\
(7) \LRL & UPR  & 20 & 0.6561 & \textbf{0.8659} & \textbf{0.6364} & 0.8129 & 0.6035 & 0.7464 \\
(8) \LRL & PRL  & 10 & 0.6369 & 0.8085 & 0.6116 & 0.7841 & 0.5844 & 0.7315 \\
(9) \LRL & PRL  & 20 & \textbf{0.6650} & 0.8405 & 0.6349 & 0.8237 & \textbf{0.6260} & 0.7689 \\
\midrule
    \multicolumn{9}{c}{\textit{Supervised}} \\
(a) DPR  & None & |$\mathcal{C}$| & 0.6297 & 0.7388 & 0.6480 & 0.8184 & -- & -- \\
(b) TCT\_ColBERT & None & |$\mathcal{C}$| & 0.7210 & \textbf{0.8864} & 0.6854 & 0.8392 & 0.5001 & 0.6527 \\
(c) MonoBERT    & BM25 & 1000 & 0.7233 & 0.8566 & 0.7218 & 0.8530 & 0.6098 & 0.7278 \\
(d) MonoELECTRA & DPR & 1000 & \textbf{0.7557} & 0.8748 & 0.\textbf{7450} & \textbf{0.8650} & -- & --\\
\bottomrule
\end{tabular}
}
\caption{
nDCG@10 and MRR@10 of different ranking pipelines on the three TREC Deep Learning tracks.
Under the ``Source'' column, ``prev.''\ describes the previous stage, thus ``None'' for first-stage retrievers.
UPR is the previous SoTA.
PRL and \LRL are approaches proposed in this paper. 
All models in the \textit{Supervised} section are trained on MS MARCO~\cite{msmarco}.
Best unsupervised and supervised results are highlighted in bold.}
\label{tab:1}
\end{table*}

\begin{table}[h]
\centering
\small
\resizebox{0.48\textwidth}{!}{ 
\begin{tabular}{l|l|l|l}
\toprule
     & Chinese & Swahili & Yoruba \\
\midrule
    \multicolumn{4}{c}{\textit{Zero-shot}} \\
(1) BM25 &  0.180 & 0.383 & 0.406 \\
(2) mContriever  & 0.212 & 0.368 & 0.187 \\
(3) Hybrid (1 \& 2) & 0.307 & 0.510 & 0.422 \\
(4) LRL  &  \textbf{0.344} & \textbf{0.542} & \textbf{0.533} \\
\midrule
    \multicolumn{4}{c}{\textit{Supervised}} \\
(a) mDPR (MS) & \textbf{0.512} & 0.299 & 0.444 \\
(b) mDPR (MS + \mrtydi) & 0.358 & \textbf{0.658} & \textbf{0.598} \\
\bottomrule
\end{tabular}
 }
\caption{
nDCG@10 on three datasets from MIRACL.
\LRL reranks top-20 results from the hybrid system in row~(3). 
Under \textit{Supervised}, mDPR (MS) is trained on MS MARCO, and mDPR (MS + \mrtydi) is further trained with \mrtydi~\cite{mdpr}. 
}
\label{tab:2}
\end{table}

We evaluate our methods on the TREC Deep Learning (DL) passage retrieval tasks:\ DL19, DL20 and DL21~\cite{dl19, dl20, dl21}.
DL19 and DL20 use the MS MARCO v1 passage corpus, which contains 8.8 million passages, and DL21 uses the MS MARCO v2 passage corpus, which contains 138 million passages.

We also evaluate on MIRACL~\citep{miracl}, which is a multilingual dataset that focuses on ad hoc retrieval across 18 different languages.
Due to cost, we conduct experiments on only three languages:\ Chinese, Swahili, and Yoruba, to verify the effectiveness of our methods on non-English (both high-resource and low-resource) languages.

Our implementation is based on the latest\footnote{At the time of our experiments.} GPT-3~\citep{gpt3} API release.
For the TREC DL datasets, we use \texttt{\small text-davinci-003} for both PRL and \LRL.
Due to the input sequence length limitation, \LRL takes 10 candidate passages (i.e., $m=10$) and truncates the length of each passage to fit the length limit when necessary.
To rerank candidate lists with a size larger than 10 (e.g., to rank the top-20 passages), we use the progressive reranking strategy described in Section~\ref{sec:rerank}.

For reranking the MIRACL datasets, we use \texttt{\small code-davinci-002}, which accepts twice the number of input tokens as \texttt{\small text-davinci-003}.
We choose a model that supports longer input lengths 
as we observe that GPT-3 tends to over-tokenize many non-English languages.

\section{Results}

\autoref{tab:1} presents results on the three TREC DL datasets, where rows~(1--9) compare different \textit{zero-shot} ranking pipelines, and rows~(a--d) report results using \textit{supervised} methods trained on the MS MARCO training set~\cite{msmarco}.
We are, of course, cognizant of potential inter-mixing of training and test data in the GPT models, so our designation of ``zero-shot'' should be taken with a grain of salt---although our work is not alone in suffering from this potential shortcoming.

Rows~(1, 2) report baselines using two first-stage retrievers, BM25 and Contriever.
Rows~(3--5) show results of different LLMs reranking top-100 candidate documents retrieved by BM25, including the current SoTA UPR, our baseline PRL, and our listwise reranker LRL.
We use BM25 as the first-stage retriever since it achieves
higher effectiveness on DL19 and DL20.
Results show that \LRL is more effective than the pointwise rerankers on all three datasets:
Take nDCG@10 as an example, on average, \LRL is around six and three points higher than UPR and PRL, respectively.

Rows~(6--9) present the results of using \LRL as an additional reranking stage to rerank the top-10/20 documents from UPR and PRL.
This setting is inspired by the multi-stage ranking approach proposed in duoBERT \cite{duobert}.
In this case, we can avoid progressive reranking over an overly long list for improved efficiency.
Results show that our approach can achieve further improvements over pointwise rerankers consistently.
For example, by reranking the top-20 PRL results, as in row~(9), \LRL improves nDCG@10 and MRR@10 by five and seven points on average.
This further supports our claim that reranking based on a list of documents indeed provides better signals than computing relevance scores for each document individually. 

Comparing the best \LRL pipelines with the supervised ranking pipelines, we are excited to see that the effectiveness of \LRL, which does not require any task-specific training, outperforms even some of the supervised dense retrievers such as DPR, which is trained on a large amount of in-domain training data, as shown in row~(a).
However, an effectiveness gap remains compared to more sophisticated supervised ranking pipelines, for example, TCT-ColBERT~\cite{tct}.

Since GPT-3 has multilingual capabilities, our proposed method can be directly applied to non-English datasets.
Table~\ref{tab:2} shows the results on three datasets from MIRACL.
In row~(3), we rerank the top-20 retrieved results from a first-stage retriever that is the hybrid of BM25 and mContriever, presented in rows~(1--2).
On the three MIRACL datasets, LRL outperforms the hybrid baseline, demonstrating the effectiveness of our \LRL reranker for multilingual retrieval.

\section{Conclusion}

In this work, we propose LRL, a listwise zero-shot reranker utilizing an LLM that adopts a simple, novel, and effective approach to reranking.
We show its improvements over the SoTA zero-shot pointwise rerankers on web search datasets and demonstrate its effectiveness across languages.
This work only begins to scratch the surface of the exciting potential of large language models for text retrieval, and we are excited about future prospects.

\section*{Acknowledgments}

This research was supported in part by the Natural Sciences and Engineering Research Council (NSERC) of Canada.

\bibliography{custom}
\bibliographystyle{acl_natbib}
\end{document}